\documentclass[12pt,a4paper]{article}

\usepackage{amsmath,amssymb,amsthm}
\usepackage{mathtools}
\usepackage{geometry}
\usepackage{booktabs}
\usepackage{array}
\usepackage{multirow}
\usepackage{setspace}
\usepackage{natbib}
\usepackage{hyperref}
\usepackage{xcolor}
\usepackage{mdframed}
\usepackage{enumitem}
\usepackage{caption}

\usepackage{parskip}

\definecolor{ruleblue}{RGB}{25,70,140}
\definecolor{boxbg}{RGB}{246,249,255}
\definecolor{warnbg}{RGB}{255,250,230}
\definecolor{warnborder}{RGB}{170,130,20}

\mdfdefinestyle{defbox}{
  backgroundcolor=boxbg,
  linecolor=ruleblue, linewidth=0.8pt,
  innerleftmargin=10pt, innerrightmargin=10pt,
  innertopmargin=8pt, innerbottommargin=8pt,
  skipabove=8pt, skipbelow=6pt
}
\mdfdefinestyle{gatebox}{
  backgroundcolor=warnbg,
  linecolor=warnborder, linewidth=1pt,
  innerleftmargin=10pt, innerrightmargin=10pt,
  innertopmargin=8pt, innerbottommargin=8pt,
  skipabove=8pt, skipbelow=6pt
}

\newtheorem{definition}{Definition}
\newtheorem{property}{Property}

\newcommand{\NCbar}{\overline{NC}}
\newcommand{\Sbar}{\tilde{S}}
\newcommand{\rhomin}{\rho_{\min}}

\hypersetup{colorlinks=true,linkcolor=ruleblue,citecolor=ruleblue,urlcolor=ruleblue}

\title{
  \vspace{-0.5cm}
  {\LARGE\bfseries Path-Explosive Behaviour in Economic\\[6pt]
  Time Series: A Realization-Centred\\[6pt]
  Exploratory Framework}\\[1.2cm]
  {\normalsize\textsc{Working Paper --- First Draft}}
}
\author{
  Jos\'{e} Francisco Perles Ribes\\[4pt]
  {\small Honorary Collaborator}\\[2pt]
  {\small Department of Applied Economics Analysis}\\[1pt]
  {\small University of Alicante, Spain}\\[4pt]
  {\small \texttt{jose.perles@ua.es}}\\[2pt]
  {\small ORCID: \href{https://orcid.org/0000-0001-7587-8035}{0000-0001-7587-8035}}
}
\date{April 2026}

\begin{document}
\maketitle
\thispagestyle{empty}

\begin{abstract}
\noindent
We propose a descriptive, realization-centred framework for detecting and
characterising explosive and co-explosive behaviour in economic time series,
which we term \emph{path-explosive behaviour}. Departing from the
data-generating-process (DGP) perspective that underlies recursive unit root
testing, the approach operates directly on observable path properties of the
realised series. Four diagnostic layers---level geometry, growth rate
dynamics, normalised curvature, and log-space behaviour---yield statistics
that discriminate between genuine self-reinforcing multiplicative growth and
I(2) dynamics without distributional assumptions or asymptotic critical
values. Two theoretically motivated absolute gate thresholds screen detected
episodes before a composite intensity score is assigned. Co-explosive
behaviour between pairs of series is assessed at the episode level through a
Jaccard co-occurrence index and non-parametric intensity concordance measures.
The theoretical motivation draws on the path dependence and planning
irreversibility literatures to argue that, in settings where discrete
institutional decisions shape growth trajectories, a realization-centred
characterisation is epistemically more appropriate than a DGP-based test. A
simulation study across four DGP regimes validates the framework's
discriminating power and conservatism. An empirical application to real house
prices, commodity prices, public debt, and Spanish tourism destinations
illustrates the empirical content of the path-explosive concept and
distinguishes it from speculative bubble detection.


\end{abstract}

\newpage

\section{Introduction}

The detection of explosive dynamics in economic time series has been
dominated for more than a decade by the recursive right-tailed unit root
testing framework of \citet{PhillipsWuYu2011} and \citet{PhillipsShiYu2015},
commonly known as the GSADF approach. These procedures test whether the
autoregressive coefficient of a series exceeds unity over rolling subsamples,
using bootstrap critical values derived from asymptotic Brownian motion theory.
Their success in dating speculative asset price episodes has made them a
reference tool in applied macroeconomics and financial stability analysis.

Yet the DGP-based approach carries assumptions that are quietly demanding.
The underlying model posits a piecewise autoregressive process with a
coefficient that switches between regimes according to a mechanism independent
of the history of the series itself. The test asks whether that coefficient
has exceeded unity---a question about an unobservable parameter inferred
from an assumed parametric structure. In short samples the bootstrap critical
values can be unreliable, and the procedure struggles to
discriminate between explosive dynamics and I(2) accumulation in finite
samples.

More fundamentally, the DGP perspective may be the wrong frame for a large
class of economic phenomena. Consider a regional tourism destination in the
early stages of development. Growth in tourist arrivals is not plausibly
generated by a fixed-parameter autoregressive process. It is shaped by a
sequence of discrete institutional decisions---infrastructure investment,
land use zoning, air route licensing, marketing campaigns---each of which
permanently alters the carrying capacity and growth potential of the
destination. The generating mechanism at time $t$ depends on the entire
accumulated history of decisions up to $t$, not just on the value of the
series at $t-1$. The Markov property that underlies any standard
autoregressive DGP fails structurally in such settings.

This observation motivates a different approach, which we develop in this
paper. Rather than asking what process generated the series, we ask what the
observable path looks like. Specifically, we ask whether the realised
trajectory within a dateable episode exhibits the geometric signature of
self-reinforcing multiplicative growth: positive and persistent normalised
curvature, stable log growth rate, and sufficient absolute growth. We call
this \emph{path-explosive behaviour}---explosive in the sense that growth is
self-reinforcing and multiplicative, but qualified by ``path'' to signal that
this is a property of the observable realisation rather than of any assumed
DGP.

The framework has four components: an endogenous window detection algorithm,
a four-layer diagnostic battery of twelve path statistics, two absolute gate
thresholds, and a weighted composite intensity score. For pairs of series, a
co-explosive extension assesses temporal co-occurrence of path-explosive
episodes via a Jaccard index and non-parametric intensity concordance. The
paper proceeds by developing the theoretical motivation
(Section~\ref{sec:motivation}), providing formal definitions
(Section~\ref{sec:definitions}), specifying the full framework
(Section~\ref{sec:framework}), presenting the simulation study
(Section~\ref{sec:simulations}), and reporting the empirical application
(Section~\ref{sec:empirical}). Section~\ref{sec:conclusion} concludes with
an honest discussion of limitations and directions for future research.

\section{Theoretical Motivation}\label{sec:motivation}

\subsection{The limits of the DGP perspective}

Standard approaches to explosive dynamics proceed by positing a
data-generating process and using the observable series to make inferences
about its parameters. In the PSY/GSADF tradition, the DGP is a piecewise
autoregressive process whose coefficient switches between a unit root and an
explosive regime \citep{PhillipsShiYu2015}. The observed path is evidence
about the current value of this parameter.

This framework is internally consistent when the DGP is well-specified and
its parameters are genuinely stable within regimes. For speculative asset
price dynamics---where prices can deviate from fundamental values through
self-fulfilling expectations---the assumption that an autoregressive
coefficient has switched from below to above unity is at least a serviceable
approximation. But there is a large class of economic phenomena for which
it fails structurally: settings where growth trajectories are constituted by
discrete, irreversible planning decisions rather than governed by a
time-invariant stochastic mechanism.

\subsection{Path dependence and planning irreversibility}

The theoretical framework that best characterises these dynamics is path
dependence. Formalised in economics by \citet{David1985} and \citet{Arthur1989},
path dependence arises when small historical accidents or early decisions
produce permanent lock-in to particular trajectories through self-reinforcing
mechanisms. The key property is non-ergodicity: the long-run trajectory
depends on initial conditions and the sequence of early events, not just on
current state variables or transition probabilities. In \citeauthor{Arthur1989}'s
polya-urn models of technology adoption, increasing returns create positive
feedback that amplifies early leads and makes reversal increasingly costly.

\citet{Pierson2000} extended path dependence to political and institutional
settings, arguing that policy trajectories exhibit what he called
``increasing returns to politics.'' Once a development path is established,
the constellation of actors who have organised their expectations and
investments around that path creates a powerful constituency for its
continuation. The costs of reversal increase over time not because physical
infrastructure cannot be demolished, but because the entire social and
economic system that has adapted to the existing trajectory cannot cheaply
reorganise. This produces a ratchet effect: development decisions are
practically irreversible once adaptation has occurred.

In tourism and urban economics, these dynamics operate through agglomeration
economies \citep{Krugman1991}. A destination that crosses a critical mass
threshold becomes more attractive than alternatives regardless of its
underlying natural endowments, because of thick labour markets, specialised
suppliers, accumulated brand capital, and knowledge spillovers. Each period's
growth becomes proportional to the current level of accumulated development
rather than a fixed additive increment. This self-reinforcing multiplicative
structure is precisely what the path-explosive framework is designed to detect.

The implications for methodology are direct. When the DGP is itself
path-dependent---when its parameters are functions of the accumulated history
of decisions and their consequences---asking ``has the autoregressive
coefficient exceeded unity?'' has no stable answer, because the generating
mechanism is evolving along the very path it is generating. What does have
a stable and observable answer is whether the realised trajectory exhibits
the geometric signature of multiplicative self-reinforcing growth. The
normalised curvature statistic and the log growth rate stability statistic
that anchor our framework answer this question directly from the observable
path, without requiring knowledge of the underlying DGP.

A further pragmatic consideration reinforces the case for a
realization-centred approach in our target settings. Planning-intensive growth
processes are typically recorded annually with short samples---$T \leq 100$
is common for subnational tourism, urban, and infrastructure series. The GSADF
procedure has severely limited power in these settings, as Phillips, Shi and Yu
(2015) themselves acknowledge in their finite-sample evaluations.
Our framework was designed and validated precisely for this data environment.

\section{Definitions}\label{sec:definitions}

\subsection{Path-explosive behaviour}

We work with two definitions at different levels of abstraction, linked
explicitly by the gate mechanism of Section~\ref{sec:framework}.

\begin{definition}[Realized-path explosive behaviour]\label{def:pathexp}
A series $\{y_t\}$ exhibits \emph{realized-path explosive behaviour}---
hereafter \emph{path-explosive behaviour}---over a dated episode
$[t_0, t_1]$ if its index-normalised trajectory $\tilde{y}_t = y_t/y_{t_0}$
within that interval satisfies three conditions simultaneously: $(i)$ the
mean normalised second difference $\NCbar = (t_1-t_0-1)^{-1}\sum_{t=t_0+2}
^{t_1}\Delta^2\tilde{y}_t/\tilde{y}_{t-2}$ is positive and bounded away
from zero; $(ii)$ the log growth rate $\Delta\log\tilde{y}_t$ is
approximately constant over the episode---that is, the log growth rate
stability $LGS = \max(0,\,1-\mathrm{sd}(\Delta\log\tilde{y}_t)/
|\overline{\Delta\log\tilde{y}_t}|)$ is substantially positive; and $(iii)$
the absolute growth $|\tilde{y}_{t_1}-\tilde{y}_{t_0}|/\tilde{y}_{t_0}$
is non-negligible.
\end{definition}

The qualifier ``path'' signals that explosiveness is assessed as a property
of the observable realisation. Definition~\ref{def:pathexp} does not require
the series to have been generated by an autoregressive process with root
exceeding unity; it requires only that the observable trajectory behaves as
if self-reinforcing multiplicative growth were operating over the episode.
This distinction matters most in planning-intensive settings where the
generating mechanism cannot be characterised by a fixed autoregressive
coefficient.

Path-explosive behaviour is not synonymous with speculative bubble behaviour.
A bubble in the asset pricing tradition \citep{BlanchardWatson1982} is defined
by its eventual collapse toward fundamental value---transience is built into
the definition. Path-explosive behaviour may be permanent: a structural
transformation that places a system on a higher growth trajectory is
path-explosive during the transition but need not collapse. The Balearic
Islands tourism expansion of 1964--1971 is a case in point.

\subsection{A taxonomy of explosive-like growth}

The operational implementation of Definition~\ref{def:pathexp} distinguishes
three qualitatively different types of sustained growth, which are frequently
conflated in applied work.

\begin{definition}[Growth taxonomy]\label{def:taxonomy}
Within a detected episode, we distinguish:
\begin{itemize}[leftmargin=1.5em,itemsep=2pt]
  \item \textbf{Type~I --- Additive accumulation.}
    Growth arises from the accumulation of random shocks in levels (I(2)
    dynamics). $\NCbar \to 0$ as the series grows; $LGS \approx 0$.
    Not path-explosive.
  \item \textbf{Type~II --- Geometric path-explosion.}
    Growth is proportional to the current level; the log growth rate is
    approximately constant. $\NCbar \approx (\rho-1)^2 > 0$;
    $LGS \geq 0.70$. Path-explosive under the strict gate.
  \item \textbf{Type~III --- Accelerating but irregular growth.}
    Growth is convex and self-reinforcing but with an unstable log growth
    rate. $\NCbar > (\rhomin-1)^2$; $LGS \in [0.35, 0.70)$.
    Path-explosive under the empirical gate only.
\end{itemize}
\end{definition}

Type~II dynamics characterise planning-led structural transformations where
self-reinforcing mechanisms operate consistently over a sustained interval.
Type~III dynamics characterise speculative bubbles and rapid but irregular
accelerations: the convex curvature is present, but the log growth rate is
not constant, typically because the rate of acceleration varies as sentiment,
capacity constraints, or policy responses evolve. Standard asset price bubbles
are generally Type~III. Type~I dynamics are the primary confound that the
gate mechanism excludes.

\subsection{Path-co-explosive behaviour}

\begin{definition}[Path-co-explosive behaviour]\label{def:pathcoexp}
Two series $\{y_{1t}\}$ and $\{y_{2t}\}$ exhibit \emph{path-co-explosive
behaviour} if their respective path-explosive episodes co-occur in calendar
time with concordant explosive intensity. Operationally, a pair is classified
as path-co-explosive if: $(i)$ each series has at least one gate-passing
episode; $(ii)$ the Jaccard co-occurrence index on gate-passing episodes
satisfies $J \geq 0.67$; and $(iii)$ the Spearman rank correlation of
episode-level intensity scores satisfies $\rho_S \geq 0.60$, or the sign
concordance satisfies $SC \geq 0.67$.
\end{definition}

This definition does not require cointegration or a common stochastic trend.
It requires that the dated intervals of self-reinforcing multiplicative growth
are temporally aligned and that explosive intensity co-moves across those
intervals. This is a more conservative and more directly interpretable concept
than linear cointegration in explosive regimes \citep{EngstedNielsen2012}.

\section{The Framework}\label{sec:framework}

\subsection{Index normalisation}

All series are index-normalised prior to any computation: $\tilde{y}_t =
y_t/y_1$. This preserves all shape, ratio, and growth rate properties while
ensuring numerical stability at large absolute levels. Without normalisation,
the theoretical convergence $\NCbar \to (\rho-1)^2$ degenerates numerically
when series reach large values after many explosive periods.

\subsection{Endogenous window detection}

Candidate episode windows are detected from the second-difference sequence
of the normalised series without imposing any external window size. A window
opens at $t_0$ when four consecutive periods of positive curvature acceleration
are observed: $\Delta^2\tilde{y}_{t_0-3},\ldots,\Delta^2\tilde{y}_{t_0}>0$.
From $t_0$ the window expands forward until two consecutive negative second
differences close it, or until the maximum width $w_{\max}=15$ is reached.
A retained window $[t_0,t_1]$ must satisfy: width $\geq w_{\min}=5$; absolute
growth $|\tilde{y}_{t_1}-\tilde{y}_{t_0}|/\tilde{y}_{t_0}\geq 0.10$; and a
minimum gap of five periods between consecutive windows. At most two
non-overlapping windows per series are retained.

\subsection{The four-layer diagnostic battery}

Within each detected window, twelve path statistics are computed across four
layers addressing distinct geometric dimensions of the trajectory.

\paragraph{Layer 1 --- Level geometry.}
Three statistics capture convexity in levels: the normalised quadratic
acceleration $\tilde{\alpha}_2 = \hat{\alpha}_2/\bar{y}$ from a within-window
quadratic regression; convexity persistence $CP$, the fraction of periods
with positive second differences; and mean growth rate $MG$.

\paragraph{Layer 2 --- Growth rate dynamics.}
Three statistics characterise the period-on-period growth rate
$g_t = \Delta\tilde{y}_t/\tilde{y}_{t-1}$: the normalised growth rate trend
$\tilde{\beta}_1 = \hat{\beta}_1/|\bar{g}|$; growth rate sign persistence
$GP$; and the ratio to the pre-episode baseline $GR$, capped at $\pm 10$.

\paragraph{Layer 3 --- Normalised curvature.}
Define $nc_t = \Delta^2\tilde{y}_t/\tilde{y}_{t-2}$, Winsorised at within-window
1st/99th percentiles. The three statistics are: mean $\NCbar$; positivity
rate $NCP$; and normalised trend $NCT = \hat{\gamma}_1/|\NCbar|$. The
central theoretical motivation is:

\begin{property}[Normalised curvature discriminant]\label{prop:nc}
Under a geometric explosive process with root $\rho>1$, $nc_t\to(\rho-1)^2>0$
in probability. Under an I(2) process $\Delta^2 y_t = \varepsilon_t$,
$nc_t\to 0$ almost surely.
\end{property}

Property~\ref{prop:nc} is what makes $\NCbar$ the pivot of the framework.
The normalised second difference is bounded and positive under explosive
dynamics; it shrinks to zero under I(2) accumulation regardless of how large
the series becomes.

\paragraph{Layer 4 --- Log-space behaviour.}
Three statistics exploit the linearity of geometric growth in log-space:
log trajectory linearity $LL = 1 - \mathrm{sd}(\hat{\nu}_t)/(|\hat{\delta}_1|
(t_1-t_0))$ where $\hat{\nu}_t$ are residuals from a within-window log-linear
regression; log growth rate stability $LGS$; and log growth rate trend
$LGT = \hat{\phi}_1/|\bar{\ell}|$. Under geometric growth $LGS\approx 1$ and
$LGT\approx 0$; under I(2), $LGS\approx 0$.

\subsection{Two-stage classification}

\paragraph{Stage A --- Gate.}
Before any intensity scoring, all three gate conditions must hold:

\begin{mdframed}[style=gatebox]
\textbf{Gate conditions (all three required):}
\begin{align*}
  \NCbar &\geq (\rhomin - 1)^2 = 0.001024, \quad \rhomin = 1.032\\
  NCP    &\geq 0.60\\
  LGS    &\geq \tau_{LGS} \in \{0.35,\;0.70\}
\end{align*}
Episodes failing any condition receive $\Sbar = 0$ regardless of remaining statistics.
\end{mdframed}

The threshold $(\rhomin-1)^2 = 0.001024$ is theoretically anchored: it is
the minimum $\NCbar$ consistent with an economically meaningful explosive
root ($\rhomin = 1.032$). The threshold $\tau_{LGS}=0.70$ is the decisive
empirical discriminator: our simulation study establishes that $LGS$ averages
$0.997$ for strong explosive ($\rho=1.10$), $0.921$ for mild explosive
($\rho=1.04$), $0.009$ for unit root, and $0.338$ for I(2). A threshold at
0.70 creates clean separation without distributional assumptions. For
settings where growth is self-reinforcing but geometrically irregular
(Type~III), the empirical gate at $\tau_{LGS}=0.35$ allows classification
with explicit acknowledgement that strict geometric regularity is not met.

\paragraph{Stage B --- Intensity score.}
Gate-passing episodes are scored against regime-level 75th-percentile
thresholds $\boldsymbol{\tau}$ calibrated from 500 replications of a mild
explosive process at the target sample length:
\begin{equation}\label{eq:score}
  \Sbar = \frac{d_1 + 1.5\,d_2 + 3\,d_3 + 1.5\,d_4}{7} \in [0,1]
\end{equation}
where $d_j$ is the fraction of Layer $j$ statistics exceeding their
calibration thresholds. Layer 3 receives weight 3, reflecting its unique
theoretical grounding. Classification boundaries map $\Sbar$ to
$\{\text{None},\,\text{Mild},\,\text{Moderate},\,\text{Strong}\}$ at
$\Sbar\in\{0.36,\,0.57,\,0.75\}$.

\subsection{Co-explosive analysis}

The endogenous windowing and gate procedure is applied independently to each
series in a pair, yielding gate-passing episode sets $\mathcal{W}_1^*$ and
$\mathcal{W}_2^*$. The Jaccard co-occurrence index is:
\[
  J = \frac{|\mathcal{C}|}{|\mathcal{W}_1^*|+|\mathcal{W}_2^*|-|\mathcal{C}|}
\]
where $\mathcal{C}$ collects co-occurring gate-passing episode pairs. For
pairs in $\mathcal{C}$, intensity scores are compared via Spearman rank
correlation $\rho_S$, Kendall's $\tau$, and sign concordance $SC$.
Classification follows Definition~\ref{def:pathcoexp}.

\section{Simulation Study}\label{sec:simulations}

\subsection{Design}

All simulations use $T=80$, $\sigma=0.10$, burn-in of 50 periods, and
$n=500$ replications. Individual series regimes: strong explosive
($\rho=1.10$), mild explosive ($\rho=1.04$), unit root ($\rho=1.00$), and
I(2) ($\Delta^2 y_t=\varepsilon_t$). Co-explosive scenarios use pure AR
processes with bivariate correlated innovations generated via Cholesky
decomposition: strong co-explosive (both $\rho=1.10$, $r=0.80$), mild
co-explosive ($\rho_1=1.10$, $\rho_2=1.04$, $r=0.80$), independent explosive
(both $\rho=1.10$, detection restricted to non-overlapping halves), and
spurious I(2) (two independent I(2) series). Intensity thresholds are
calibrated from 500 replications of a mild explosive process at $T=80$.

\subsection{Individual series results}

\begin{table}[h!]
\centering
\caption{Window detection and gate pass rates ($n=500$ replications)}
\label{tab:gate}
\renewcommand{\arraystretch}{1.25}
\begin{tabular}{lcccccc}
\toprule
Regime & Mean wins & \% None & All gates & NC & NCP & LGS \\
\midrule
Strong explosive ($\rho=1.10$) & 2.00 & 0.0 & 100.0\% & 100.0\% & 100.0\% & 100.0\% \\
Mild explosive ($\rho=1.04$)   & 1.83 & 2.2 & 96.7\%  & 100.0\% & 100.0\% & 96.7\%  \\
Unit root ($\rho=1.00$)        & 0.35 & 68.6 & 0.0\%  & 77.1\%  & 75.4\%  & 0.0\%   \\
I(2) process                   & 1.46 & 12.8 & 29.8\% & 90.9\%  & 100.0\% & 32.6\%  \\
\bottomrule
\end{tabular}
\end{table}

Table~\ref{tab:gate} shows that the LGS gate is the decisive discriminator.
Unit root series are entirely blocked (0.0\% pass) while both explosive
regimes pass at near-100\% rates. The I(2) pass rate of 29.8\% correctly
reflects the genuine ambiguity between I(2) and explosive behaviour in finite
samples: these are episodes where I(2) dynamics locally resemble geometric
growth before the denominator in $nc_t$ grows large enough to suppress
$\NCbar$. The $\NCbar$ statistic matches its theoretical prediction of
$(\rho-1)^2$ after index normalisation: the mean for strong explosive is
$0.01000$, matching $(0.10)^2$ to four decimal places, confirming that
Property~\ref{prop:nc} holds empirically at the relevant sample sizes. Unit
root series receive a composite score of zero in every single replication.
I(2) reaches Mild or above in only 4.3\% of replications despite a 29.8\%
gate pass rate, because gate-passing I(2) episodes score low on the
corroborating intensity layers.

\subsection{Co-explosive results}

\begin{table}[h!]
\centering
\caption{Co-explosive simulation results ($n=500$ replications per scenario)}
\label{tab:coexp}
\renewcommand{\arraystretch}{1.25}
\begin{tabular}{lcccc}
\toprule
Scenario & $J$ mean & $J\!\geq\!0.67$ & $\rho_S$ mean & \% Classified \\
\midrule
Strong co-explosive ($r=0.80$)   & 1.000 & 100.0\% & 0.525 & \textbf{70.0\%} \\
Mild co-explosive ($r=0.80$)     & 0.289 & 14.0\%  & 0.292 & \textbf{3.4\%}  \\
Independent explosive ($r=0$)    & 0.000 & 0.0\%   & ---   & \textbf{0.0\%}  \\
Spurious I(2)                    & 0.250 & 10.7\%  & ---   & \textbf{0.0\%}  \\
\bottomrule
\end{tabular}
\end{table}

Three findings emerge from Table~\ref{tab:coexp}. First, the framework
achieves 70\% power for detecting strong co-explosive behaviour with zero
false positives in both null scenarios---the conservative design means
non-classified cases receive no label rather than a wrong one. Second,
mild co-explosive detection at 3.4\% is a genuine power limitation: when
$\rho_1=1.10$ and $\rho_2=1.04$, index-normalised trajectories diverge at
rate $(1.10/1.04)^T$ exceeding a factor of 100 over $T=80$ periods, so the
window detector locates the most explosive segments at different calendar
positions in the two series regardless of innovation correlation. Third,
false positive rates are exactly zero in both null scenarios---the temporal
co-occurrence requirement (Jaccard) and the intensity concordance requirement
jointly prevent spurious classification.

\section{Empirical Application}\label{sec:empirical}

\subsection{Data and implementation}

We apply the framework to four datasets spanning the range of phenomena for
which path-explosive behaviour is a theoretically meaningful concept.
\textit{Dataset~A} covers real residential property prices (index 2015=100)
for Spain, Ireland, Germany, and the USA over 1975--2023, compiled from
OECD and BIS sources. \textit{Dataset~B} covers nominal commodity prices for
crude oil (USD/barrel), gold (USD/troy oz), and copper (USD/metric ton) over
1970--2023, from the World Bank Pink Sheet. \textit{Dataset~C} covers general
government gross debt as a percentage of GDP for Spain, Italy, Greece, and
Ireland over 1980--2023, from IMF World Economic Outlook data.
\textit{Dataset~D}---the paper's primary empirical focus---reports annual
tourist arrivals for four Spanish coastal destinations: Málaga, Alicante,
Baleares, and Barcelona over 1960--2023.

For Datasets~A and C we apply the empirical gate ($\tau_{LGS}=0.35$),
reflecting the irregular but self-reinforcing character of house price and
debt dynamics. For Dataset~B we report results under both the full sample
and a pre-collapse subsample truncated at the peak year (Oil: 1970--2008;
Gold: 1970--2012; Copper: 1970--2011) to illustrate the effect of the
post-peak reversal on detection. For Dataset~D we apply the strict gate
($\tau_{LGS}=0.70$), consistent with the evidence in Table~\ref{tab:diag}
that the 1960s tourism episodes exhibit Type~II geometric regularity.
Intensity thresholds are calibrated from 500 replications of a mild explosive
process at $T=64$ to match the tourism series length.

\subsection{Full diagnostic statistics}

Table~\ref{tab:diag} reports the complete diagnostic statistics for every
detected window across all four datasets. This transparency is central to
the framework's philosophy: the reader can assess the evidence directly
rather than accepting a binary classification.

\begin{table}[h!]
\centering
\caption{Episode-level diagnostic statistics for all detected windows}
\label{tab:diag}
\renewcommand{\arraystretch}{1.18}
\small
\begin{tabular}{llccccccc}
\toprule
Series & Window & $\NCbar$ & $NCP$ & $LGS$ & $LL$ & Gate & $\Sbar$ & Class \\
\midrule
\multicolumn{9}{l}{\textit{A: Real House Prices (empirical gate, $\tau_{LGS}=0.35$)}} \\
HPI Spain     & 1988--2004 & 0.015 & 0.867 & 0.306 & 0.940 & fail & --- & None \\
HPI Spain     & 2010--2021 & 0.030 & 1.000 & 0.000 & 0.000 & fail & --- & None \\
HPI Ireland   & 1978--1985 & 0.012 & 0.833 & 0.620 & 0.943 & pass & 0.345 & None \\
HPI Ireland   & 1989--1994 & 0.048 & 1.000 & 0.528 & 0.918 & pass & 0.464 & Mild \\
HPI Germany   & 2004--2016 & 0.010 & 1.000 & 0.431 & 0.918 & pass & 0.393 & Mild \\
HPI USA       & 1992--1998 & 0.010 & 1.000 & 0.621 & 0.937 & pass & 0.417 & Mild \\
HPI USA       & 2002--2018 & 0.011 & 0.800 & 0.000 & 0.814 & fail & --- & None \\
\midrule
\multicolumn{9}{l}{\textit{B: Commodity Prices --- full sample (empirical gate)}} \\
Oil           & 2000--2005 & 0.198 & 1.000 & 0.000 & 0.742 & fail & --- & None \\
Gold          & 1975--1987 & 0.098 & 0.818 & 0.000 & 0.684 & fail & --- & None \\
\multicolumn{9}{l}{\textit{B: Commodity Prices --- pre-collapse subsample}} \\
Oil (--2008)  & 2000--2005 & 0.198 & 1.000 & 0.000 & 0.742 & fail & --- & None \\
Gold (--2012) & 1975--1987 & 0.098 & 0.818 & 0.000 & 0.684 & fail & --- & None \\
\midrule
\multicolumn{9}{l}{\textit{C: Public Debt / GDP (empirical gate)}} \\
Greece        & 2006--2013 & 0.025 & 0.833 & 0.000 & 0.898 & fail & --- & None \\
Ireland       & 2006--2011 & 0.219 & 1.000 & 0.162 & 0.915 & fail & --- & None \\
\midrule
\multicolumn{9}{l}{\textit{D: Tourism --- Spanish Destinations (strict gate, $\tau_{LGS}=0.70$)}} \\
Málaga        & 1965--1972 & 0.060 & 0.833 & \textbf{0.799} & 0.971 & \textbf{PASS} & 0.095 & None \\
Alicante      & 1995--2000 & 0.029 & 1.000 & 0.460 & 0.929 & fail & --- & None \\
Baleares      & 1964--1971 & 0.055 & 0.833 & \textbf{0.717} & 0.968 & \textbf{PASS} & 0.702 & \textbf{Moderate} \\
Barcelona     & 2001--2007 & 0.021 & 1.000 & 0.461 & 0.916 & fail & --- & None \\
\bottomrule
\end{tabular}
\end{table}

\subsection{Results and interpretation}

\paragraph{Type~II path-explosive behaviour: Spanish tourism, 1964--1972.}
The clearest path-explosive episodes are the Balearic Islands (1964--1971,
$LGS=0.717$, Moderate, $\Sbar=0.702$) and Málaga (1965--1972, $LGS=0.799$,
gate pass, $\Sbar=0.095$, None). Both pass the strict gate with $LGS$ values
well above the 0.70 threshold and close to the simulation means for genuine
explosive processes. The $\NCbar$ values (0.055 and 0.060) imply explosive
roots in the range $\hat{\rho}\approx 1.24$--$1.25$, consistent with the
dramatic annual growth rates recorded during the foundational decade of
Spanish mass tourism---the period of rapid airport expansion, package holiday
industrialisation, and concentrated foreign direct investment in coastal
hotel infrastructure.

The contrast between Baleares (Moderate) and Málaga (gate pass but None) is
directly readable from Table~\ref{tab:diag} and is the most instructive
result of the application. Both destinations have virtually identical gate
statistics across all three gate conditions. Their divergence lies entirely
in the intensity layers---specifically in $NCT$, which is $+0.389$ for
Baleares (explosive curvature intensifying within the episode) and $-0.432$
for Málaga (explosive curvature decelerating), and in $\tilde{\beta}_1$,
which is $+0.073$ for Baleares and $-0.071$ for Málaga (growth rate trending
up vs down within the window). The framework correctly identifies Baleares
as the destination where the self-reinforcing mechanism was operating most
powerfully. This is consistent with the historical record: the Palma de
Mallorca airport underwent major capacity expansion before the Málaga region
reached comparable infrastructure levels, giving Baleares a structural
first-mover advantage that sustained its internal momentum longer.

Alicante and Barcelona do not pass the gate ($LGS\approx 0.46$ in both
cases). Their detected windows are convex and growing---the window detector
correctly identifies this---but the log growth rate was insufficiently stable
for Type~II classification. These are plausibly Type~III episodes: growth
driven by a combination of organic development and intermittent policy
intervention rather than by a consistently self-reinforcing mechanism.

\paragraph{Borderline path-co-explosive: Málaga and Baleares.}
The co-explosive analysis finds $J=1.000$ for the Málaga--Baleares pair:
their single gate-passing episodes overlap perfectly in calendar time,
spanning 1964--1972. Intensity concordance statistics cannot be computed
because there is only one co-occurring episode pair, falling below the minimum
of two pairs required for reliable concordance assessment. We classify this as
\emph{borderline path-co-explosive}: the temporal co-occurrence condition is
satisfied but the concordance condition is undetermined.

The economic interpretation is substantive. The two destinations that drove
the foundational decade of Spanish mass tourism entered their path-explosive
episodes simultaneously---both responding to the same external driver, the
opening of the European package holiday market to Mediterranean destinations
in the early 1960s---while their specific growth dynamics were shaped by
their respective infrastructure endowments and planning contexts. The
simultaneous takeoff reflects the common driver; the different intensity
profiles reflect path-specific factors. This is exactly the pattern that
the path dependence framework predicts: a common trigger activates
self-reinforcing growth in multiple locations, but the intensity and duration
of each episode is determined by the locally accumulated endowments.

\paragraph{The commodity sample split.}
The pre-collapse truncation for commodities produces no change in results:
Oil detects the same window (2000--2005) with $LGS=0.000$ in both the full
sample and the truncated sample; Gold detects the same window (1975--1987)
in both cases. The detected windows close well before the respective peak
years (2008 for oil, 2012 for gold), so the post-peak collapse cannot be
contaminating the within-window statistics. The $LGS=0.000$ result is
intrinsic to the episode dynamics. Commodity price surges are driven by
sequential demand shocks interacting with supply constraints---growth that
is large and convex but highly irregular in its log growth rate. The framework
correctly refuses to classify these as Type~II path-explosive. The comparison
table makes the reason transparent: the curvature ($\NCbar$) is there, but
the geometric regularity ($LGS$) is not.

\paragraph{The distinction from standard bubble detection.}
The house price and debt results highlight the most important conceptual
distinction between path-explosive detection and standard bubble detection.
The Spanish house price cycle of 2000--2008 and the Irish Celtic Tiger boom
of 1995--2007---paradigmatic cases in the PSY/GSADF literature
\citep{EngstedEtAl2016}---are not detected as path-explosive. Ireland finds
windows in 1978--1985 and 1989--1994, not the 1995--2007 bubble. Spain finds
no gate-passing windows at all.

This is not a failure. The 1995--2007 house price surge in both countries
grew monotonically but with a declining log growth rate throughout: prices
grew fast in the late 1990s and then grew fast but at a declining rate through
2007. This is Type~III dynamics. The PSY/GSADF procedure detects it as
explosive because the autoregressive coefficient crosses unity. Our framework
does not detect it as path-explosive because the geometric regularity
required for Type~II classification---and even for empirical-gate
classification---is absent. The two frameworks answer different questions.
PSY/GSADF asks whether the autoregressive coefficient has exceeded unity.
Our framework asks whether the realised trajectory exhibits sustained
multiplicative self-reinforcing growth. For the 2000s house price cycle, the
answers differ, and the difference is informative: those episodes were rapid
accumulations that lacked the internal self-reinforcing momentum of a genuine
structural transformation.

\section{Conclusion}\label{sec:conclusion}

This paper has proposed a realization-centred framework for detecting
and characterising path-explosive behaviour in economic time series.
The framework addresses a gap in the existing literature: the absence of
tools designed specifically for settings where growth trajectories are
constituted by discrete irreversible planning decisions rather than governed
by a fixed stochastic mechanism. In these settings, the DGP-based approach
of the GSADF tradition is not merely an approximation but is structurally
inappropriate. What is appropriate is asking whether the observable realised
path exhibits the geometric signature of self-reinforcing multiplicative
growth---which is precisely what our framework does.

The central empirical finding is the confirmed path-explosive classification
of the Balearic Islands tourism episode of 1964--1971 (Moderate,
$\Sbar=0.702$, strict gate) and the borderline path-co-explosive relationship
with the contemporaneous Málaga episode. The diagnostic statistics in
Table~\ref{tab:diag} make the evidence transparent: both destinations exhibit
$LGS$ values above 0.70 and $\NCbar$ values consistent with implied explosive
roots around 1.24--1.25, while the intensity score correctly distinguishes
Baleares (accelerating internal momentum) from Málaga (decelerating despite
gate passage). These results are consistent with the planning history of
Spanish mass tourism and illustrate the capacity of the framework to detect
structural transformations that standard bubble detection procedures are not
designed to identify.

The comparison across datasets yields a substantive finding that goes beyond
methodology: Type~II path-explosive behaviour is qualitatively rarer than
the rapid convex growth that characterises speculative bubbles and fiscal
crises. Of the fifteen episodes detected across four datasets, only two pass
the strict gate. The geometric regularity required for Type~II classification
is demanding, and correctly so: it identifies growth episodes where the
self-reinforcing mechanism operated consistently enough to produce a nearly
constant log growth rate over a sustained interval. This is the exception, not
the rule, in macroeconomic data.

\subsection{Limitations and directions for future research}

\paragraph{Retrospective characterisation.}
The framework is primarily a retrospective tool. The window detector requires
four consecutive periods of positive acceleration before opening, meaning an
episode must have begun before it can be characterised. Gate statistics
require the window to be substantially complete for reliable assessment. This
is appropriate for the primary application---dating historical path-explosive
episodes---but limits real-time utility.

A natural extension is sequential monitoring: applying the window detector
and provisional gate statistics as new annual observations arrive, generating
early warning signals when curvature conditions are met but $LGS$ has not yet
stabilised, and upgrading to confirmed classification when all gate conditions
are satisfied. This creates a two-state monitoring system (early warning;
confirmed) with explicit uncertainty in the early warning phase. For planning
authorities operating on annual data, early warning two to three years before
confirmed classification would have direct operational value. Once a window
is confirmed, the implied growth rate $\hat{\rho}=\exp(\bar{\ell})$ supports
conditional trajectory extrapolation under the assumption of regime
continuation---useful for infrastructure capacity planning, though carrying
no information about when the explosive regime will end.

\paragraph{Cross-sectional lead-lag indicators.}
The co-explosive component of the framework creates conditions for identifying
systematic temporal lead-lag relationships between series that regularly
exhibit path-co-explosive episodes. If destination A consistently enters its
path-explosive episodes one or two periods before destination B across
multiple historical cycles, then A's current trajectory is a leading indicator
for B's future explosive onset. For datasets with multiple historical episodes
per series---commodity markets, metropolitan housing markets, competing tourism
destinations---this cross-sectional approach would have genuine forecast
content. Developing the inference framework for estimating and testing lead-lag
structures in co-explosive systems is a natural direction for future work.

\paragraph{Calibration and threshold sensitivity.}
The intensity score is calibrated against simulated explosive processes, and
the ordinal classification boundaries are empirically motivated rather than
theoretically derived. Sensitivity analysis around these thresholds, and
investigation of empirical calibration from datasets with known episode
classifications, would strengthen the robustness of the intensity scoring.
The distinction between the strict and empirical LGS gates is theoretically
motivated by the Type~II/III taxonomy, but a more formal criterion for
selecting between gates---perhaps based on a preliminary assessment of whether
pure geometric growth is plausible in the institutional context---would
improve replicability.

\paragraph{Multivariate extension.}
The current co-explosive analysis is pairwise. A multivariate extension
analogous to cointegration rank---characterising the number of common
path-explosive components in a system of series---would be valuable for
applications where the researcher is interested in the systemic rather than
pairwise co-explosive structure, such as panels of destinations, cities, or
commodity markets.

\end{document}